# EXPERIMENTAL EVALUATION OF BIRD STRIKES IN URBAN AIR MOBILITY

Aditya Devta[1], Isabel C. Metz[1], and Sophie F. Armanini[2]

[1]Institute of Flight Guidance, German Aerospace Center DLR
[2]TUM School of Engineering and Design, Technical University of Munich

**Abstract**

Since the Wright brothers demonstrated the first powered, sustained and controlled flight in 1903, the airspace has been shared between birds and humans. Novel aircraft and advanced mobility concepts such as Urban Air Mobility are emerging in full swing. In that concept, a safe and efficient aviation transportation system will use highly automated aircraft that will transport passengers or cargo at low altitudes within and between metropolitan regions. To accomplish these missions, new types of aircraft which are sometimes known as air taxis are being developed. A successful integration of these aircraft into existing airspace is complicated and needs to take into account various aspects. One of these is the risk of wildlife strikes in general and bird strikes in particular. While bird strike constitutes a risk to any type of aircraft, the risk is predicted to be higher in case of air taxis. The proposed operational cruising altitude of air taxis is lower resulting in higher probability of collision as these are the altitudes where birds typically fly. Additionally, air taxis are smaller in size and have lower certification requirements compared to conventional aircraft. As a result, the severity of damaging bird strikes is higher. To assess the risk and formulate suitable regulations, an extensive analysis is required providing more quantitative insight into the bird strike challenge. Therefore, a theoretical model of bird strike to quantify the impact force exerted due to a strike by considering different bird and aircraft related parameters was developed previously. This paper aims to validate this theoretical model experimentally. It presents a methodology for implementing an experimental setup, allowing for the theoretical impact force model to be fully validated. A test matrix containing seven test cases, nine test scenarios and 135 iterations is formulated to conduct the bird strike experiment and the influencing parameters are considered for theoretical model verification. The paper closes with the presentation of the experimental results for validating the theoretical model which indicate 92.89 % conformance of experimental results with the theoretical model.

## NOMENCLATURE

| | | | |
|---|---|---|---|
| $\Delta E_{kinetic}$ | Kinetic energy transfer | $\rho_{material}$ | Density of the impacted material |
| m | Mass of the bird | $V_{projectile}$ | Volume of the projectile |
| $v_{bird}$ | Speed of the bird | $V_{material}$ | Volume of impacted material |
| $v_{aircraft}$ | Speed of the aircraft | A | Surface area |
| $\rho_{aircraft}$ | Density of the aircraft | $d_{cylinder}$ | Depth of penetration of cylindrical bird |
| d | Depth of penetration | $d_{ellipsoid}$ | Depth of penetration of ellipsoidal bird |
| r | Radius of the bird | v | Impact velocity |
| $\rho_{bird}$ | Density of the bird | a | Acceleration of the projectile |
| l | Length of the bird | $C_d$ | Drag co-efficient |
| v | Volume of the bird | k | Constant |
| $\theta$ | Angle of impact | t | Time |
| $\rho_{projectile}$ | Density of the projectile | h | Drop height |

## 1   INTRODUCTION

To alleviate the increase in traffic in the metropolitan environment, advanced transportation concepts are being proposed and UAM(Urban Air Mobility) is one of them [1]. UAM is a futuristic mobility concept which involves development of eVTOL(Electric Vertical Take Off and Landing) aircraft for transporting commuters or freight within and around the urban environment with operational altitudes below 1219 m (4000 ft) [2]. With the introduction of these eVTOL aircraft, which will be referred to as air taxi in the rest of this paper, the low altitude airspace is going to witness an increase in air traffic [3]. At the same time, long-term statistics suggest that 88 % of the reported collisions between birds and conventional air traffic have occurred below this altitude (92% up to 3500 ft, 94% up to 4500 ft) [4]. This indicates a high probability of bird strikes for these newly developed air taxis. Moreover, air taxis are expected to be smaller and fly at a lower cruising speed in the range of 77.6 - 89 m/s [5]. As a result, bird speeds considered to be insignificant in the case of conventional aircraft can substantially impart higher kinetic energy and thus higher impact force to air taxis. Since air taxis are subjected to less stringent certification requirements as compared to commercial airliners [6], a high likelihood of damaging strikes is to be expected. This suggests a significantly increased overall risk of bird strike representing a safety hazard to both aircraft and birds. Hence, it is vital to quantify the consequences of potential bird strikes to analyze the impact on the safe integration of UAM traffic in the existing urban airspace. This problem was partly addressed in [7], which proposed a theoretical impact force model that evaluates the effects of collision in terms of kinetic energy and impact force. Furthermore, the influence of various bird- and aircraft-related parameters on bird strike severity was determined. Based on the obtained results, current certification requirements [8] were considered and suggestions were made for potential adjustments to these requirements for the UAM case. With the results forming the foundation for the here presented study, their major outcomes are described in the next section.

### 1.1 Current certification requirements and recommendations

The existing proposal by the EASA(European Union Aviation Safety Agency) outlines the certification requirement for single bird strikes, which entails withstanding a collision with a 1 kg bird at critical cruise speed [8]. In the case of bird flocks, the requirement is to endure a 0.45 kg bird strike at critical cruise speed. Additionally, the windshield positioned in front of occupants and the corresponding supporting structures must be able to withstand bird impacts without any penetration when the aircraft reaches maximum speeds of 25 m/s [8].

The study [7] considered different influencing parameters namely bird mass, aircraft speed, bird length, bird speed, angle of impact, aircraft material density and penetration depth. The major effect on the impact force was observed from aircraft speed, bird speed and bird mass, in the given order. Therefore, a more precise estimation of impact force could be achieved by incorporating factors such as bird velocity and aircraft skin density. This, in turn, could facilitate the establishment of more precise certification requirements for air taxis. Additionally, the results indicated that reducing the impact angle can significantly reduce the force of impact. Thus, it might be advantageous to employ more curved designs for the aircraft's fuselage, along with implementing supplementary systems that allow for swift adjustments of the impact angle.

After exhibiting the recommendations to the certification specifications based on the analytical impact force model, certain constraints and shortcomings of the model along with the necessity of experimental validation are illustrated below.

1) Ensuring accuracy: The theoretical impact force model is based on assumptions and simplifications that may not hold true in real-world scenarios. Validating the model through practical experiments helps to verify its accuracy and improve its predictive capability.
2) Identifying limitations: Practical experiments can reveal limitations or weaknesses in the theoretical model that may have been overlooked or not considered. These limitations can then be addressed and improved upon, leading to a more robust and accurate model.
3) Enhancing understanding: Validating theoretical models with practical experiments can help gain a deeper understanding of the bird strike scenario. The experiments can provide insights that may not have been anticipated or predicted by the model, leading to new discoveries and ideas.
4) Enhancing credibility: Validating a theoretical model with practical experiments will enhance its credibility and acceptance.

For these reasons, the goal of this paper is to validate the theoretical impact force model experimentally by developing an experimental setup representing collisions between air taxis and birds. Taking the underlying factors influencing the collision between air taxis and birds into account, four aspects of the experimental setup which will represent the influencing parameters have been identified : launching mechanism, bird projectile, test specimen and sensors.

The structure of this paper is as follows : Section 2 describes the modelling of the theoretical impact force and the step by step approach for developing the experimental setup of bird strikes before presenting the various aspects of the validation setup. Eventually, the different test cases and the test matrix for performing experimental tests are defined. Section 3 explains the key results obtained by executing the experimental tests and presents a comparison of these results with reference values obtained from the theoretical model. A critical discussion of the results and their implications is provided in section 4. In section 5, the primary findings are outlined and key conclusions are summarized, along with suggestions for future research.

## 2 METHODOLOGY

The goal of this research was to validate the theoretical impact force model presented in [7] of bird strikes through experimental tests. The model quantifies the impact force and kinetic energy exerted due to a strike by considering different bird and aircraft related parameters. The modelling is presented in the next section.

### 2.1 Modelling of the Theoretical Impact Force

This section provides the key equation of the theoretical impact force model as introduced in [7]. It was assumed that the bird is approaching the air taxi at an impact angle $\theta$. If the collision is head on, then $\theta$ is equal to $90°$. Impact force $F$ is generally defined as [9]

$$F = \frac{\Delta E_{kinetic}}{d} \cdot sin\theta \tag{1}$$

The basic kinetic energy equation for a projectile according to the laws of motion is as follows.

$$\Delta E_{kinetic} = \frac{1}{2} \cdot m \cdot v^2 \quad (2)$$

For the impact scenario, total speed can be expressed as

$$v = v_{bird} \cdot sin\theta + v_{aircraft} \quad (3)$$

Hence, substituting Equation 3 in Equation 2,

$$\Delta E_{kinetic} = \frac{1}{2} \cdot m \cdot (v_{bird} \cdot sin\theta + v_{aircraft})^2 \quad (4)$$

In order to reduce complexity, the bird was modelled as a right circular cylinder. The height of the cylinder was assumed to be equal to the length of each bird, while the radius was calculated for each bird based on its volume, which in turn was obtained from its known density and mass. Penetration depth was modelled as a function of bird length $l$, bird density $\rho_{bird}$, aircraft density $\rho_{aircraft}$, bird speed $v_{bird}$ and aircraft speed $v_{aircraft}$. Based on conservation of linear momentum, penetration depth for impacting birds is represented in the Equation 5. Please refer to [7] for detailed modelling.

$$\boxed{d_{cylinder} = l \cdot \frac{\rho_{bird}}{\rho_{aircraft}} \cdot \frac{v_{bird} \cdot sin\theta + v_{aircraft}}{v_{aircraft}}} \quad (5)$$

Substituting values of Equation 5 and Equation 4 in Equation 1, the impact force $F$ can be modelled as follows.

$$\boxed{F = \frac{\frac{1}{2} \cdot m \cdot \rho_{aircraft} \cdot v_{aircraft} \cdot (v_{bird} \cdot sin\theta + v_{aircraft})}{l \cdot \rho_{bird}} \cdot sin\theta} \quad (6)$$

After modelling the impact force mentioned in Equation 6, the next section outlines a methodology for developing an experimental setup to validate the theoretical impact force model

### 2.2 Experimental Setup

In order to accomplish the mentioned goal of this research, an experimental setup representing a collision between air taxis and birds was developed to validate the theoretical model with experimental results. However, experimental validation poses its own challenge in terms of physical representation of the system and level of measurement error, this section provides a step by step approach for building the experimental setup representing a bird strike. The proposed experimental model quantifies the impact force exerted due to a bird strike and compares the achieved results with the theoretical baseline which was obtained from Equation 6. The bird species and their key characteristics relevant for this experimental study were obtained from [10], [11]. Their summary is found in [7]. This bird data served as a basis for developing the experimental model of a bird strike. Sample birds were chosen to cover a wide range of sizes and masses, allowing for general conclusions to be drawn. Subsequently, it was crucial to identify the underlying variables influencing the bird strike in order to quantify their influence on the resulting impact force. The theoretical model defined in Equation 6 represents the underlying factors of a bird strike. The experimental model had to represent these factors while simulating a bird strike. Hence, based on these factors, four aspects of the experimental model were identified representing these factors. The summary is provided in Table 1 [12].

| Variable | Symbol | Representative Aspect in the Model |
|---|---|---|
| Speed of the bird | $v_{bird}$ | Launching mechanism |
| Speed of the aircraft | $v_{aircraft}$ | |
| Mass of the bird | $m$ | Bird projectile |
| Density of the bird | $\rho_{bird}$ | |
| Length of the bird | $l$ | |
| Radius of the bird | $r$ | |
| Shape of the bird | – | |
| Density of the aircraft | $\rho_{aircraft}$ | Test specimen |
| Angle of Impact | $\theta$ | |
| Kinetic Energy | $E_{kinetic}$ | Sensors |
| Impact Force | $F$ | |

TABLE 1: Underlying factors influencing the bird strike (based on [7])

After identifying the four aspects of the model presented in Table 1, the different alternatives of the launching mechanism, the bird projectile, the test specimen and the sensors were considered in this research. They were judged against technical and operational requirements. In the following sections, the details regarding selected alternative and its function in the experimental model is presented for all the four aspects.

### 2.2.1 Launching Mechanism

The function of the launching mechanism was to bring the bird projectile into motion and launch it against the test specimen. For the launching mechanism, different alternatives were considered in this research and were judged against the technical requirements such as ability of launching birds with varying masses and velocities, handling bird projectiles with different densities and operational requirements such as having low complexity and cost. The evaluation suggested that a drop-weight method was the most suitable solution for the launching mechanism as it was the only mechanism to satisfy all the requirements. A drop-weight method is a technique in which a projectile of known mechanical properties falls onto a test specimen from a specified drop-height under the influence of gravity. The impact velocity of the projectile is varied by setting the appropriate drop-height. A simple schematic of the test setup showing the drop-height, bird element and aircraft element is shown in Figure 1. Neglecting the air resistance, the impact velocity for a freely falling projectile is expressed in Equation 7 [13].

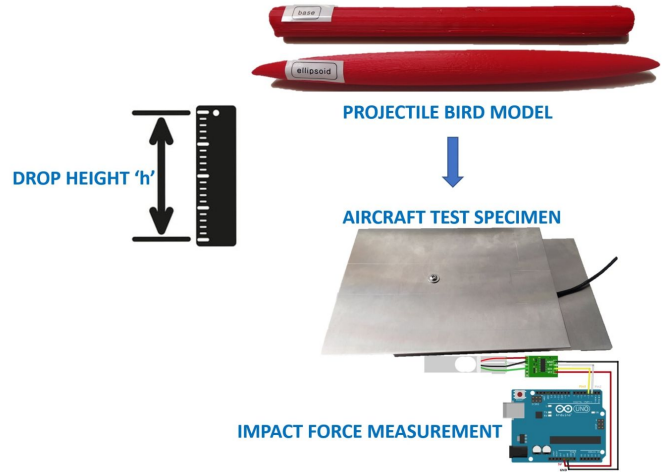

Fig. 1: Schematic diagram of the test setup showing the drop-height

$$v = \sqrt{2 \cdot g \cdot h} \qquad (7)$$

If the air resistance is not neglected, then the motion of the projectile will be influenced by acceleration due to gravity and the aerodynamic drag. The governing equation for the impact velocity as a function of time is presented in Equation 8 [14] assuming that there were no external forces acting on the projectile apart from air resistance and gravity and there is only vertical component of velocity present in the free fall motion.

$$v = \sqrt{\frac{2 \cdot m \cdot g}{\rho \cdot C_d \cdot A}} \cdot \tanh\left(\sqrt{\frac{g \cdot \rho \cdot C_d \cdot A}{2m}} \cdot t\right) \qquad (8)$$

In addition to this, the impact velocity in the experimental model was a combination of two underlying variables namely the aircraft speed and the bird speed. Hence, the impact velocity can also be expressed as presented in Equation 9.

$$v = v_{bird} + v_{aircraft} \qquad (9)$$

Moreover, the expected cruising speed of air taxis is between 77.16 m/s (150 knots) and 102.88 m/s (200 knots) [5]. Therefore, the mid value of the cruise speed was selected for calculations in this paper which results in 90 m/s (175 knots). The values of bird speeds can be found from the bird data available in the literature [10], [11] (see [7] for summary). The next section addresses the calculations of the required drop-height for performing the experimental tests.

### 2.2.2 Calculation of the required drop-height

Drop-height for this experiment was the height from which the projectile bird was released to impact against the test specimen. After analysing Equations 7 and 9, it was inferred that the required drop-height depends on the magnitude of aircraft speed and bird speed. For simplicity, the aerodynamic drag was neglected, as in Equation 7. Combining Equations 7 and 9, the expression for drop-height was obtained as follows.

$$h = \frac{v^2}{2 \cdot g} \tag{10}$$

$$h = \frac{(v_{bird} + v_{aircraft})^2}{2 \cdot g} \tag{11}$$

Using Equation 11, the drop-height was calculated for all the bird species studied in [7] and it is presented in Table 2.

| Species | Original impact velocity (m/s) | Original drop-height (metres) | Scaled impact velocity (m/s) | Scaled drop-height (metres) |
|---|---|---|---|---|
| Common Grackle | 103.41 | 535 | 6.89 | 2.4 |
| Starling | 112.35 | 631 | 7.49 | 2.8 |
| House Sparrow | 102.77 | 528 | 6.85 | 2.3 |
| Mallard | 119.06 | 709 | 7.94 | 3.1 |
| Turkey Vulture | 116.82 | 708 | 7.79 | 3.0 |
| Laughing Gull | 96.70 | 467 | 6.44 | 2.0 |
| Bald Eagle | 110.12 | 606 | 7.34 | 2.7 |
| Canada Goose | 107.88 | 582 | 7.19 | 2.6 |
| Rock Dove | 126.11 | 795 | 8.40 | 3.5 |
| Ring-billed Gull | 107.88 | 582 | 7.19 | 2.6 |
| Herring Gull | 107.88 | 582 | 7.19 | 2.6 |

TABLE 2: Original and scaled impact velocity and corresponding drop-height of the projectile bird

As observed in Table 2, drop heights ranging between 467 m and 795 m would be required to achieve the actual bird and aircraft speeds. These values were not achievable in the available test facility infrastructure because of height constraints. To mitigate this problem, the impact velocity was scaled down by 1:15 to reduce the drop-height into the feasible range of the current test facility. The scaled down impact velocity and the corresponding drop-height are also presented in Table 2.

It is important to note that because of scaling down the impact velocity, the drop-weight method was not able to reproduce realistic bird and aircraft speeds. However, the requirement of this research was to have the ability to vary different underlying parameters and analyse the impact force measurements. It was not necessary to recreate the actual aircraft and bird speeds as the goal of this study was to validate the theoretical impact force model. The consequences of scaling down are further discussed in section 5. The next section explains the details of the bird projectile.

### 2.2.3 Bird Projectile

The function of the bird projectile was to reproduce the motion and geometry of a real bird and impact the test specimen. To emulate the real bird, cylindrical projectiles were developed. The selection of the bird projectile material was made on the basis of a study conducted by Wilbeck and Rand [15]. They concluded that avian creatures can be accurately modelled using a material whose density is slightly greater than the density of water which is equal to 1000 $kg/m^3$ [15]. Consequently, different alternatives of materials satisfying the density criteria were investigated. These were Gelatin, Rubber, Polymer Clay, ABS(Acrylonitrile Butadiene Styrene) and HIPS(High Impact Polystyrene Sheet). These options were evaluated against requirements such as having material density similar to bird flesh and sustaining the impact by not breaking up. The evaluation indicated that both ABS and HIPS were able to comply with all the criteria. Additionally, these materials were available as 3D printing filaments. The main advantage of the 3D printing filaments is that the projectile can be modelled in any CAD(Computer Aided Design) software in the required shape, size, mass and density, enabling the tailoring according to the requirements of the experiment. ABS was selected to model the projectile bird because of its availability. The next section describes the manufacturing cycle of the projectile which are used in the experiment.

### 2.2.4 Manufacturing of the Bird Projectile

The bird projectiles were manufactured using the 3D printing process. FreeCAD [16] was chosen to design the CAD models of the projectiles. To be consistent with the theoretical model to be validated, the projectiles were modelled as right circular cylinders and ellipsoid shapes. After defining the shape of the projectile, the next step was to specify its geometric dimensions. The dimensions were calculated as follows.

$$V = \pi r^2 l = \frac{m}{\rho_{bird}} \tag{12}$$

$$r = \sqrt{\frac{m}{\rho_{bird} \cdot \pi \cdot l}} \tag{13}$$

Thereby, the length of the cylindrical projectile equals the bird length as obtained from [10], [11]. The resulting dimensions can be found in Table 3.

| Species | Cylinder radius (metres) | Cylinder height (metres) |
|---|---|---|
| Common Grackle | 0.01 | 0.31 |
| Starling | 0.01 | 0.22 |
| House Sparrow | 0.007 | 0.16 |
| Mallard | 0.03 | 0.57 |
| Turkey Vulture | 0.03 | 0.72 |
| Laughing Gull | 0.02 | 0.43 |
| Bald Eagle | 0.06 | 0.90 |
| Canada Goose | 0.05 | 0.92 |
| Rock Dove | 0.02 | 0.33 |
| Ring-billed Gull | 0.02 | 0.48 |
| Herring Gull | 0.03 | 0.66 |

TABLE 3: Resulting geometrical specifications of the projectile bird

The experiment was performed exemplarily for the bird species of Starling. The projectile's effect was dependent on the five parameters mass, density, length, radius and shape (cf. Table 1). Therefore, five projectiles were modelled and then manufactured by varying each of the underlying variables once. The geometric dimensions of the projectiles are shown in Table 4. The variation in density was achieved by different amounts of material infill. The values of 15% and 40% were selected since they were the minimum and maximum available presets for material infill in the 3D printer.

| Cylindrical Projectiles | | | | |
|---|---|---|---|---|
| Projectile SN(Serial Number) | Cylinder radius (metres) | Cylinder height (metres) | Material Infill amount (%) | Varying Factor |
| 1 | 0.01 | 0.22 | 15 | Base model |
| 2 | 0.01 | 0.22 | **40** | Bird density & bird mass (Infill) |
| 3 | **0.005** | 0.22 | 15 | Bird radius (& bird mass) |
| 4 | 0.01 | **0.15** | 15 | Bird length (& bird mass) |
| Ellipsoidal Projectile | | | | |
| Projectile Serial Number (SN) | Principal length a (metres) | Lateral dimensions b & c (metres) | Material Infill amount (%) | Varying Factor |
| 5 | 0.01 | 0.22 | 15 | Bird shape |

TABLE 4: List of projectiles and their respective dimensions

Using these geometrical specifications, the 3D models and prints of the bird projectiles are shown in Figure 2.

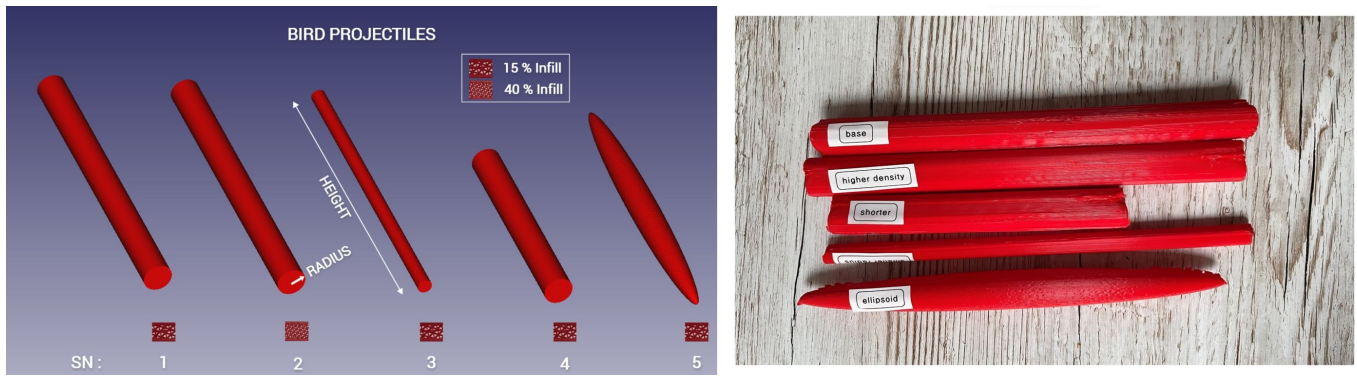

Fig. 2: CAD models and 3D prints of the bird projectiles

The next section presents the test specimen used in the bird strike experiment.

### 2.2.5 Test Specimen

In the bird strike experiments, the test specimen was a piece of material representing the aircraft skin or the aircraft structure. The test specimen will be subjected to the impact force generated due to the bird strike. The material of the test specimen should represent common aerospace structural materials of air taxis. According to eVTOL manufacturers, Aluminium-2024-T3 and CFRP(Carbon Fibres Reinforced Plastic) are widely used structural materials for air taxis [17]. Hence, Aluminium-2024-T3 and CFRP were judged against the defined requirements for the test specimen. Since both materials were suitable, they both were used for the test specimen. This had the added benefit of being able to vary the density of the aircraft by changing the material of the specimen. To represent average fuselage skin thickness of air taxis [18], the thickness of the specimen was equal to $0.002\,m$. To fit into the experimental setup, the length of the specimen was $0.2\,m$ and the width $0.15\,m$. The procured test specimens are depicted in Figure 3.

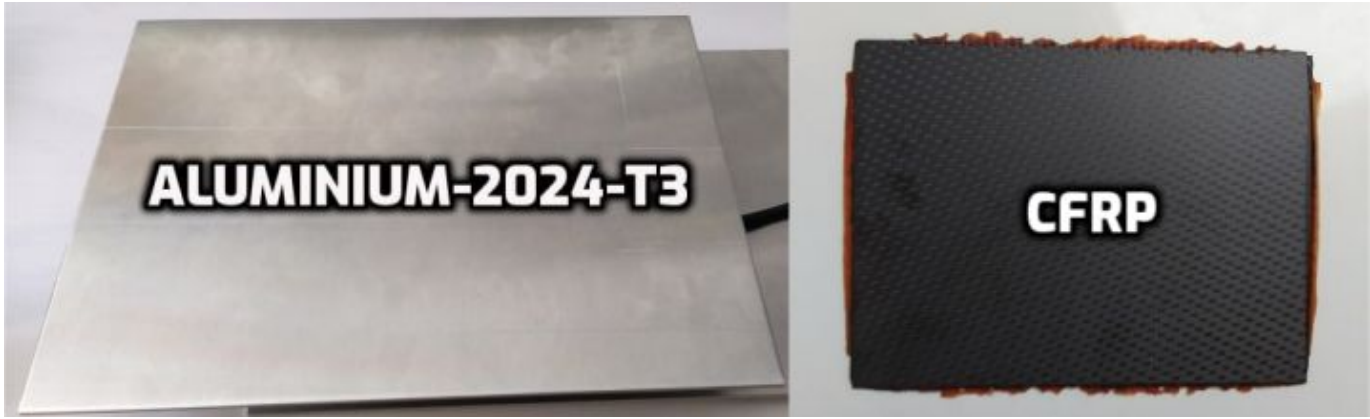

Fig. 3: Test specimen for the bird strike experiment. Left: Aluminium-2024-T3; right: CFRP

The next section introduces the sensors used in the experiment, their functions and the data acquisition system.

### 2.2.6 Sensors

In the context of this experiment, the function of the sensor was to measure the impact force exerted by the bird projectile on the test specimen. Different force sensor alternatives of FSR(Force Sensing Resistor), piezoelectric force sensor and load cell were evaluated. Since the load cell fulfilled the set requirements best, this option was selected. Load cells convert a force such as tension, compression, pressure, or torque into voltage signals that can be measured and standardized [19]. The procured load cell for this experiment is illustrated in Figure 4.

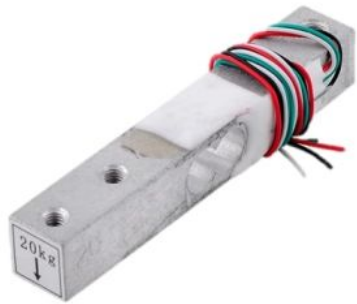

Fig. 4: Procured load cell force sensor

After selecting the appropriate sensor for measuring the impact force, the next step involved installing the sensor in the experimental model and developing a data acquisition system. These steps are presented in the next section.

### 2.2.7 Data Acquisition and Data Processing

Data acquisition was required to sample and digitize the raw electrical signals generated by the load cell force sensor. Data processing was required to process the digital signal, so that it can be deciphered by a computer and a human. Since the change was infinitesimal, an amplifier was required. The data acquisition and processing system used in the bird strike experiment consisted of a HX711 module, Arduino microcontroller board and LCD(Liquid Crystal Display) display. It is depicted in Figure 5.

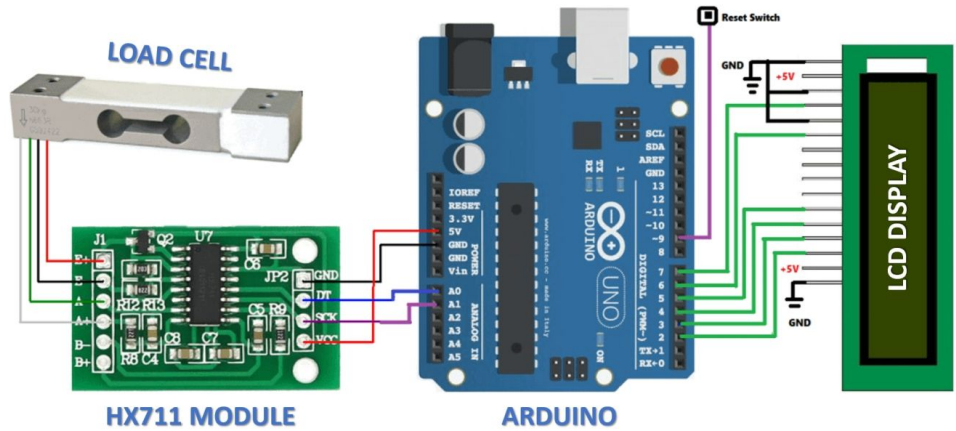

Fig. 5: Data acquisition and processing system in the bird strike experiment

The HX711 module is a breakout board that works both as an amplifier as well as an ADC(Analog to Digital Converter) [20]. The main function of the HX711 module was to amplify the low-voltage sensor signals generated by the load cell and then convert the amplified signals into digital numeric values. In this research, a precision 24 bit HX711 amplifier was used to accomplish the mentioned task. The output of the HX711 was fed to an Arduino microcontroller board.

Arduino board [1] is a microcontroller [21] for open source prototyping projects. In this research, the Arduino collects the sensor data sent by the HX711 module, and calibrates, processes and forwards it to a LCD unit and a MATLAB based software for data acquisition and visualization. On the LCD, the impact force was displayed. The LCD unit is depicted in Figure 6.

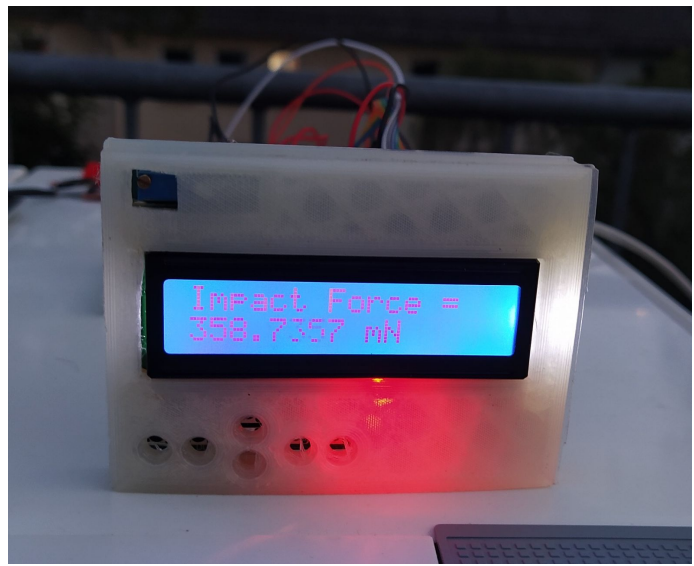

Fig. 6: 16x2 Liquid Crystal Display unit

The several functions of the MATLAB interface developed for this research were to receive the serial data sent by the Arduino, process the data, generate live serial data plots of the impact force values, log the impact force values and visualize them using a GUI(Graphical User Interface). The snapshot of the MATLAB GUI is shown in Figure 7.

---

[1]Following common practice, the Arduino microcontroller board will be referred as just Arduino

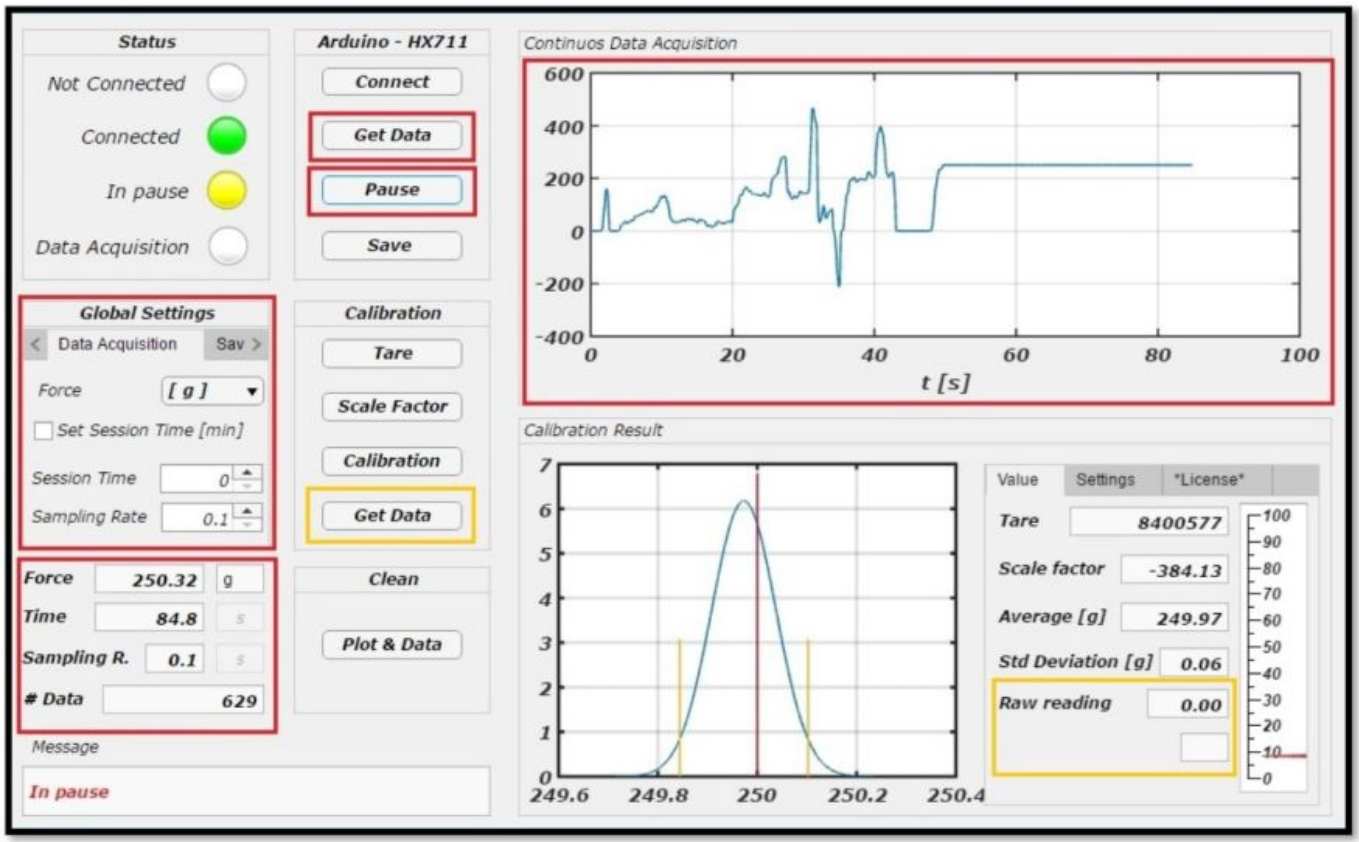

Fig. 7: MATLAB interface for data acquisition and visualization

At this point, the specific details of all four aspects of the bird strike experimental model namely the launching mechanism, the bird projectile, the test specimen and the sensors were accumulated. Using this information, the bird strike experimental model was assembled and the final assembly is illustrated in Figure 8. The main data flow between the elements is shown in Figure 5.

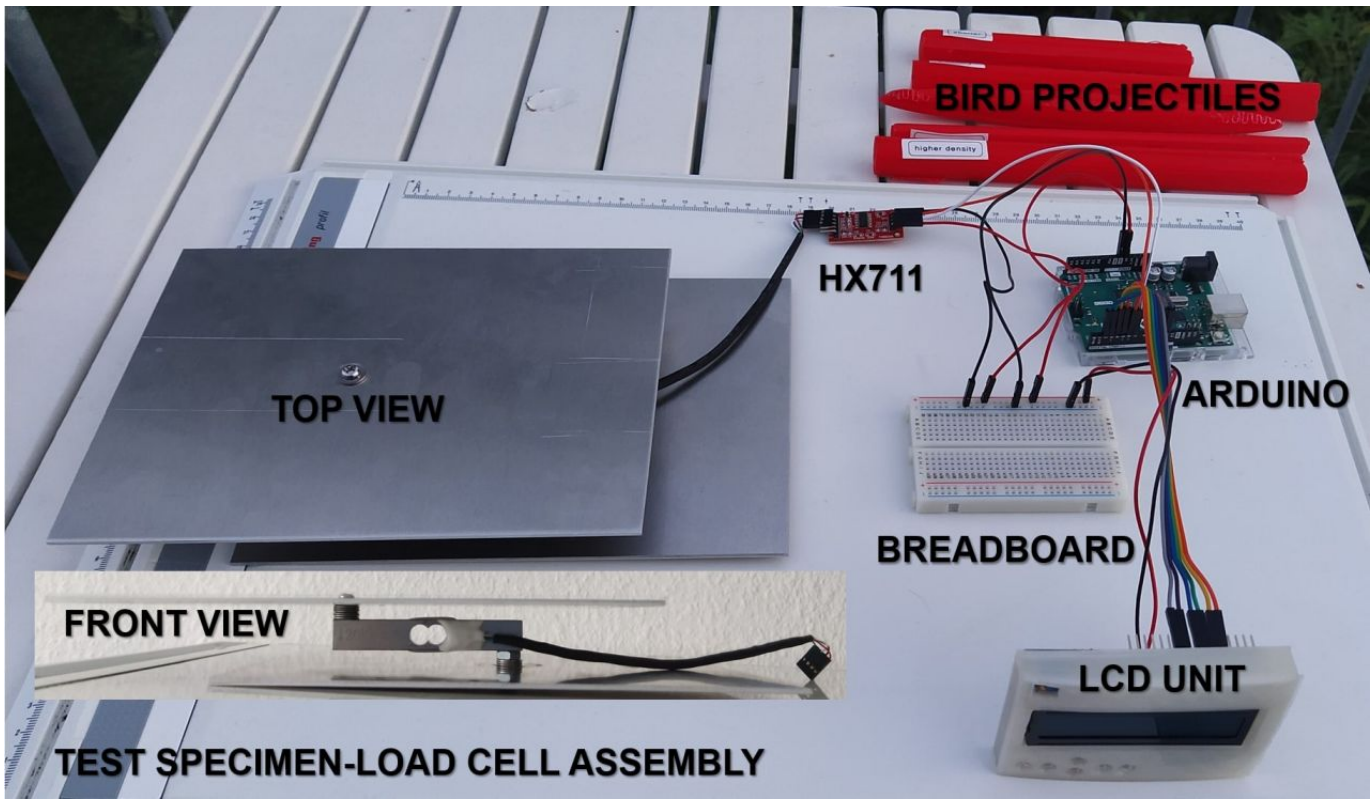

Fig. 8: Bird strike experimental test setup

After assembling the bird strike model, it was essential to quantify the actual impact velocity obtained in the particular test iteration, as it may differ from the theoretical value obtained from bird data because of factors such as aerodynamic drag and wind resistance. Moreover, according to the impact force model presented in Equation 6, the impact velocity was required as an input parameter for the theoretical estimation of impact force. The methodology employed in this research for impact velocity measurement was as follows. A high-speed tracking camera was used to record the video of the bird projectile impacting the test specimen from the specified drop-height. Subsequently, the recorded video was reconstructed using a video analysis and motion tracking tool called Tracker. Tracker is an open source software for performing manual and automated object tracking with position, velocity and acceleration overlays and data [22]. After providing the drop-height and the characteristic properties of the projectile such as mass and density as input parameters, Tracker uses the Equation 8 and quantifies the impact velocity of the projectile for the particular test iteration by using the recorded time difference between drop time & impact time. This time difference was calculated by mapping the time stamp of the video, frames per second and the provided drop-height. The snapshot of the video analysis performed in Tracker for this experiment along with the Tracker user interface is depicted in Figure 9. The left part of the image was the main video view of Tracker and the test setup is shown. The top right part shows the motion tracking plot of the projectile by displaying the drop-height and time difference for each video frame. In the bottom right, the plot data is visualized in tabular format.

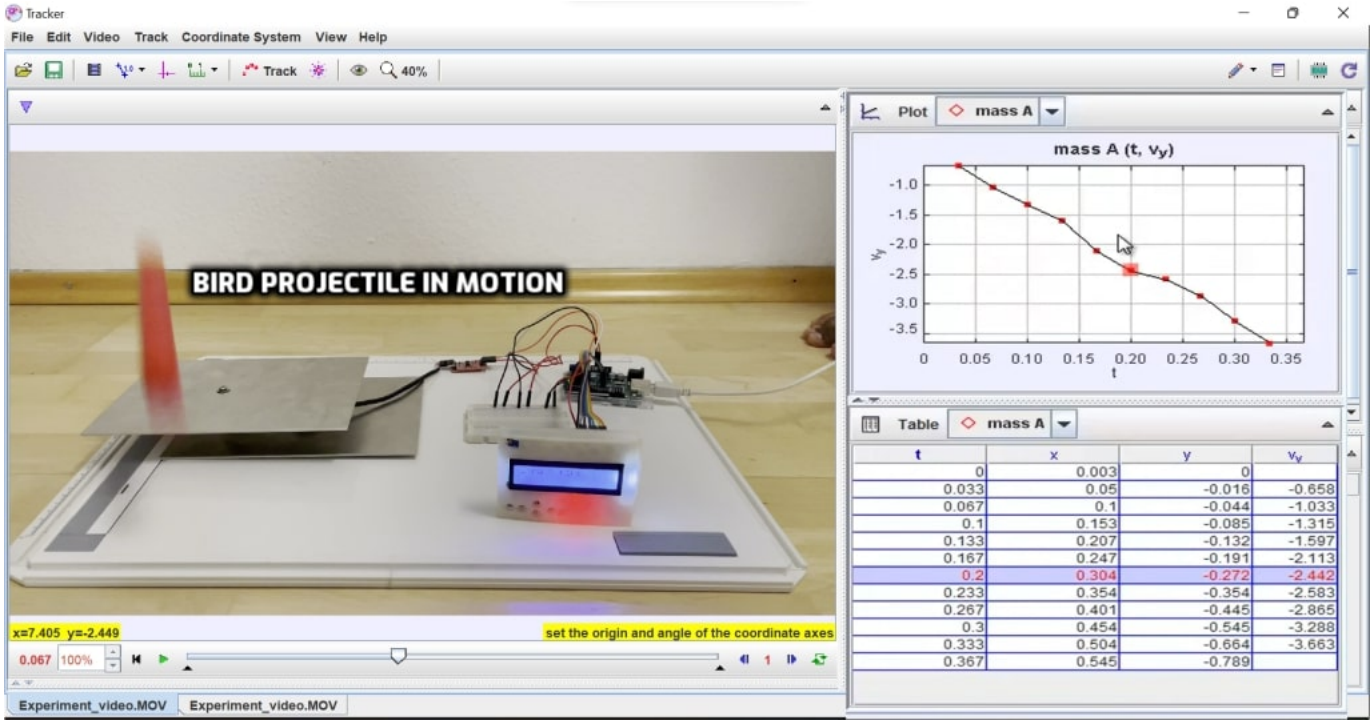

Fig. 9: Video analysis performed in motion tracking tool Tracker for impact velocity measurement

The resulting impact velocity was then used as an input parameter in Equation 6 to estimate the theoretical impact force. Moving further, the next section formulates the different test cases of the experiment and finally presents a test matrix.

## 3 Validation

To validate the model in different scenarios, seven test cases were formulated to test the effect of adjusting different impact parameters. In each of the test cases, the bird projectile was dropped from a specified drop-height and impacted against the test specimen. To gather enough data points, 15 iterations were performed in each of the test cases. These seven test cases are explained in the next sections. If not specified otherwise, the drop height was 2.8 m, corresponding to the scaled impact velocity of Starlings equal to 7.5 m/s and the angle of impact was perpendicular to the test specimen of Aluminium, corresponding to 90%. These conditions were referred to as standard experimental conditions with the scenarios with projectile SN1 serving as baseline.

### 3.1 Test case 1 : Validating the influence of bird mass

To analyze the influence of bird mass, the two bird projectiles SN1 and SN3 (cf. Table 4) only differing in their mass were used.

### 3.2 Test case 2 : Validating the influence of impact velocity

This test case was divided into two parts.

#### 3.2.1 Test case 2.1 : Validating the influence of bird and aircraft speed

In addition to the standard drop-height of 2.8 m, the projectile was also dropped from 2 m to vary the impact velocity, and the impact force was measured for each set of drops.

#### 3.2.2 Test case 2.2 : Validating the influence of bird speed for zero aircraft speed

The impact force model for stationary aircraft can be obtained by substituting the value of $v_{aircraft}$ equal to 0 in Equation 8.

$$F = \frac{\frac{1}{2} \cdot m \cdot v_{bird}^2 \cdot \rho_{aircraft} \cdot sin^3\theta}{l \cdot \rho_{bird}} \quad (14)$$

The projectile was dropped from 1.5 m which corresponds to the impact velocity at zero aircraft speed for Starlings.(see Table 2).

### 3.3 Test case 3 : Validating the influence of bird density

To validate the influence of bird density, two bird projectiles SN1 and SN2 (cf. Table 4) varying in material density were utilized.

### 3.4 Test case 4 : Validating the influence of bird length

To evaluate the influence of bird length, projectile SN4 having distinct bird length was dropped additionally to projectile SN1.

### 3.5 Test case 5 : Validating the influence of angle of impact

To quantify the effect of the impact angle, the standard setup was varied by tilting the test specimen by $50°$ and compared to the baseline scenario with $90°$.

### 3.6 Test case 6 : Validating the influence of aircraft density

In this test case, the experiment was performed under standard conditions for both test specimens, Aluminium and CFRP to assess the impact of aircraft density.

### 3.7 Test case 7 : Validating the influence of bird shape

In order to validate the effect of bird shape, two projectiles SN1 and SN5 (cf. Table 4) having different shapes were used.

### 3.8 Formulation of Test Matrix

After defining all the test scenarios, a test matrix was formulated and it is presented in Table 5. For some of the test cases, multiple scenarios were performed to compare the test condition to the baseline scenario. Each scenario was repeated 15 times to ensure repeatability and a sufficient number of data points. In total, 135 iterations were performed. The next section presents the key results obtained from the experiments according to the test cases and test matrix described in section 3.

| Test case | Description | Influencing variable | Experimental specifications | | | | | Constant variables |
|---|---|---|---|---|---|---|---|---|
| | | | Scenario | Projectile SN | Drop-height (m) | Impact velocity (m/s) | Angle of impact (°) | Test specimen |
| 1 | Validating the influence of bird mass | bird mass | Baseline | 1 | 2.8 | 7.49 | 90° | Aluminium | $\rho_{bird}$, $v_{bird}$, $\rho_{aircraft}$, $v_{aircraft}$, $\theta$ |
| | | | 1 | 3 | 2.8 | 7.49 | 90° | Aluminium | |
| 2 | Validating the influence of impact velocity | bird speed & aircraft speed | Baseline | 1 | 2.8 | 7.49 | 90° | Aluminium | m, r, l, $\rho_{bird}$, $\rho_{aircraft}$, $\theta$ |
| | | | 2.1 | 1 | 2.0 | 6.44 | 90° | Aluminium | |
| | | | 2.2 | 1 | 1.5 | 5.47 | 90° | Aluminium | |
| 3 | Validating the influence of bird density and bird mass | bird density & bird mass | Baseline | 1 | 2.8 | 7.49 | 90° | Aluminium | r, l, $v_{bird}$, $v_{aircraft}$, $\rho_{aircraft}$, $\theta$ |
| | | | 3 | 2 | 2.8 | 7.49 | 90° | Aluminium | |
| 4 | Validating the influence of bird length and bird mass | bird length & bird mass | Baseline | 1 | 2.8 | 7.49 | 90° | Aluminium | r, $v_{bird}$, $\rho_{bird}$, $v_{aircraft}$, $\rho_{aircraft}$, $\theta$ |
| | | | 4 | 4 | 2.8 | 7.49 | 90° | Aluminium | |
| 5 | Validating the influence of angle of impact | angle of impact | Baseline | 1 | 2.8 | 7.49 | 90° | Aluminium | m, r, l, $v_{bird}$, $\rho_{bird}$, $v_{aircraft}$, $\rho_{aircraft}$ |
| | | | 5 | 1 | 2.8 | 7.49 | 50° | Aluminium | |
| 6 | Validating the influence of aircraft density | aircraft density | Baseline | 1 | 2.8 | 7.49 | 90° | Aluminium | m, r, l, $v_{bird}$, $\rho_{bird}$, $v_{aircraft}$, $\theta$ |
| | | | 6 | 1 | 2.8 | 7.49 | 90° | CFRP | |
| 7 | Validating the influence of bird shape | bird shape | Baseline | 1 | 2.8 | 7.49 | 90° | Aluminium | m, r, l, $v_{bird}$, $\rho_{bird}$, $v_{aircraft}$, $\rho_{aircraft}$, $\theta$ |
| | | | 7 | 5 | 2.8 | 7.49 | 90° | Aluminium | |

TABLE 5: Test matrix of the bird strike experiment

## 4 RESULTS

According to the test matrix illustrated in Table 5, results were obtained for seven test cases and nine test scenarios which covered variations in all the influencing parameters bird mass, bird velocity, aircraft velocity, bird density, bird length, angle of impact, aircraft density and bird shape. Figures 10 - 16 illustrate the results of the baseline scenario along with the results of the individual test scenarios and compare the resulting impact force obtained through theoretical estimation and through the experimental tests. This allowed to validate the theoretical model for variations in the parameters in the individual test cases and to evaluate the effect of the influencing variables.

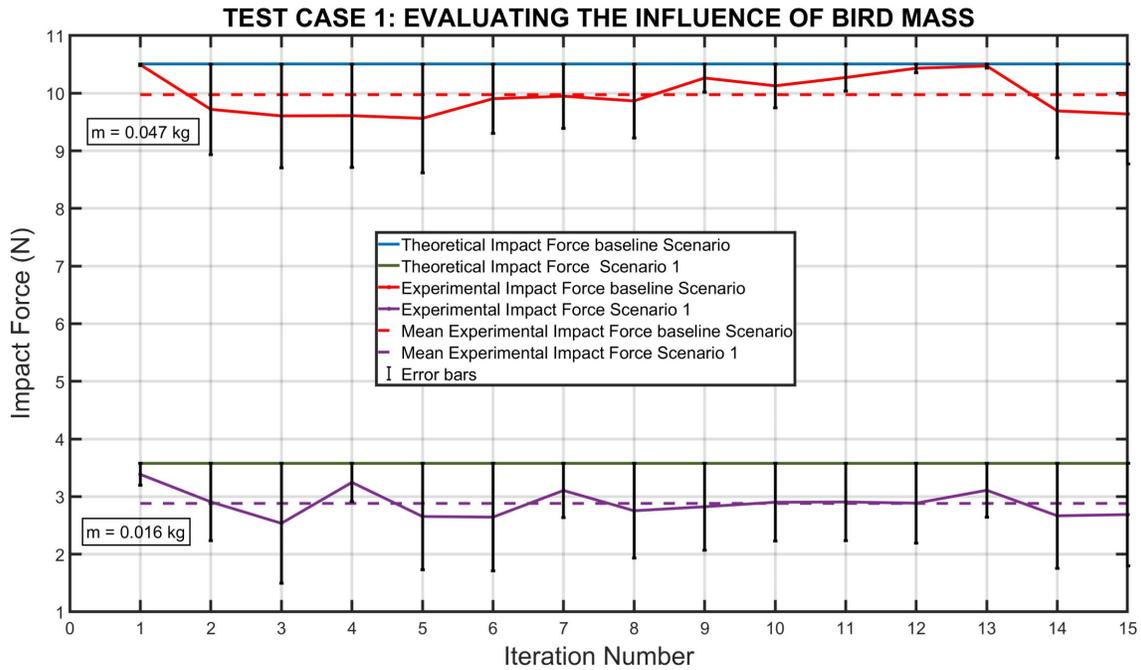

Fig. 10: Influence of bird mass (Test Case 1)

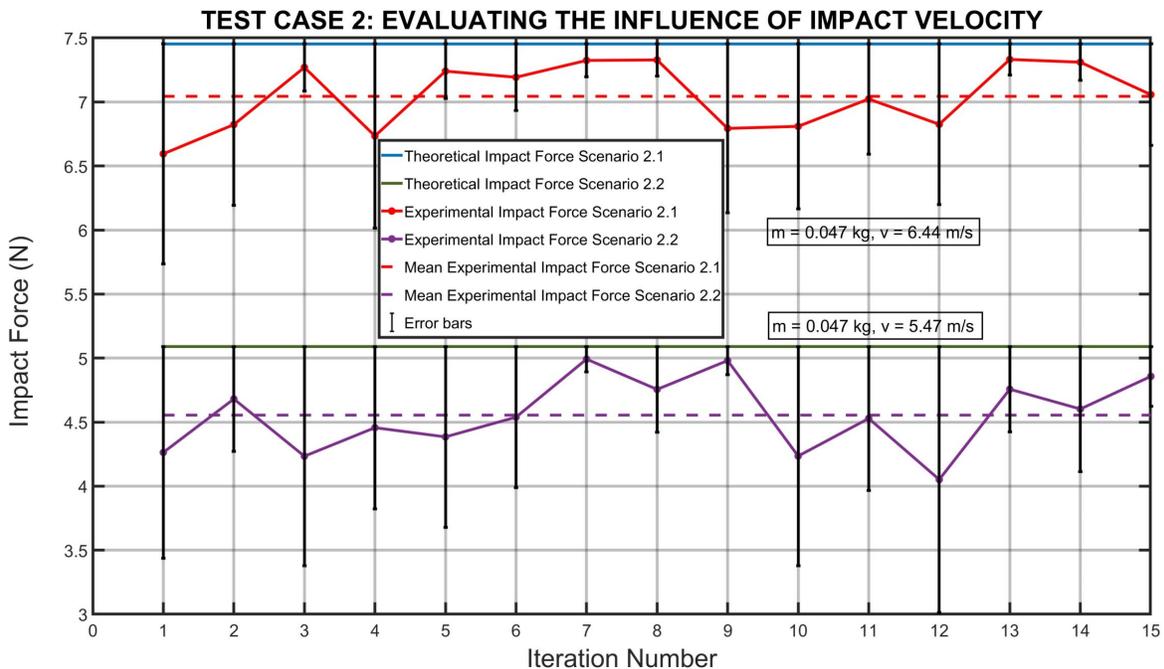

Fig. 11: Influence of impact velocity (Test Case 2)

Figure 10 shows the impact of bird mass. The lighter bird creates a lower impact as would be expected, reducing the impact force by 65 % in theory. The average reduction in the practical experiment amounted to 71 %.

Figure 11 displays the effect of impact velocity, which consists of the velocities of both the bird and the aircraft. The results depict that by increasing the impact velocity of the projectile by 17 %, the projectile exerts 50 % and 55 % more impact force theoretically and experimentally, respectively.

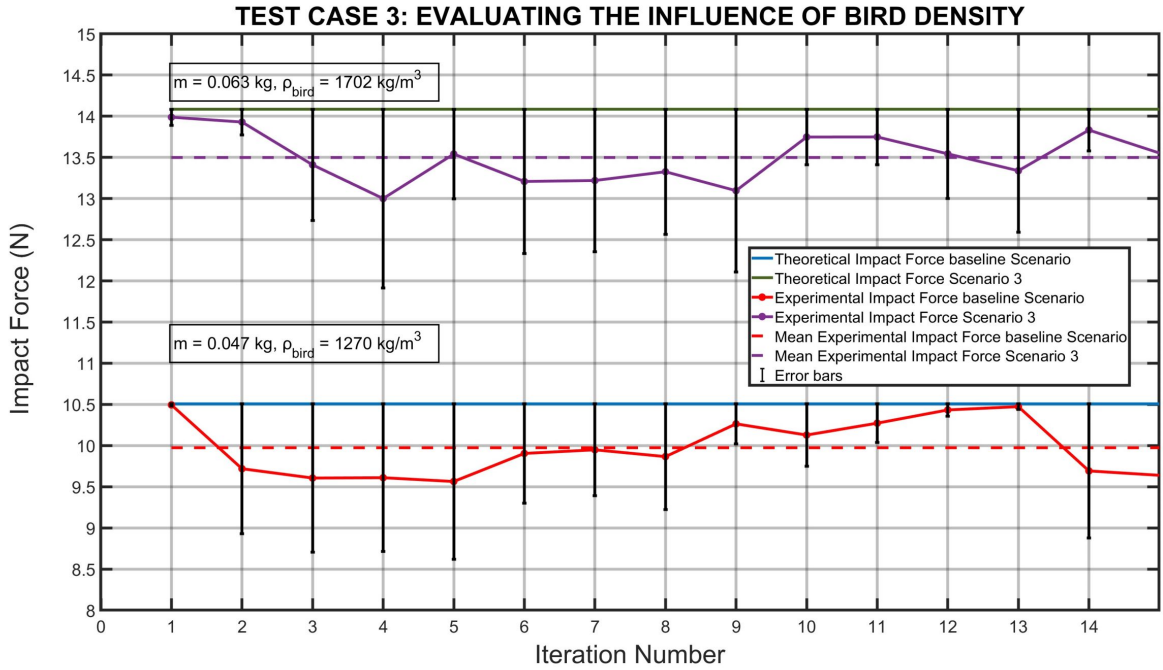

Fig. 12: Influence of bird density (Test Case 3)

Figure 12 illustrates the influence of bird density. The results demonstrate that the bird projectile, with a 34 % higher density, imparts a 40 % higher impact force in theory. Experimentally, the mean increase in impact force was measured to be 36 %.

The impact of bird length is shown in Figure 13. The results depict that 31 % reduction in bird length contributed to 2 % reduction in impact force, both experimentally and theoretically.

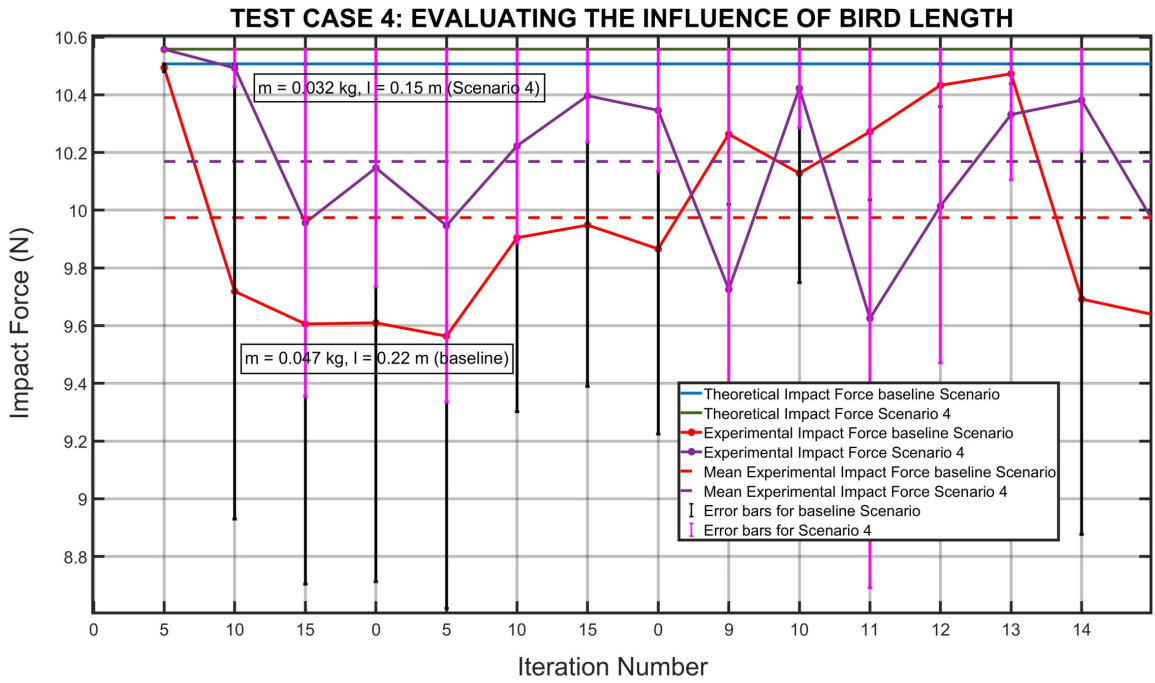

Fig. 13: Influence of bird length (Test Case 4)

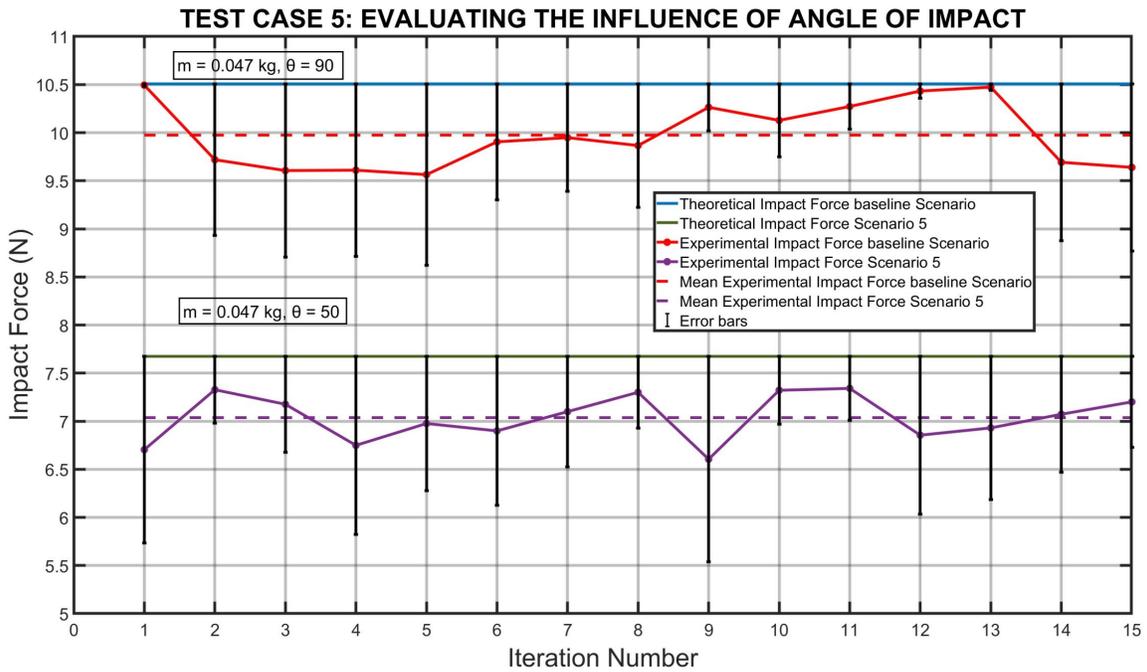

Fig. 14: Influence of impact angle (Test Case 5)

In Figure 14, the effect of impact angle is depicted. By tilting the test specimen to 50° (as compared to 90° in the standard setup), both theoretical and experimental results show a reduction in impact force of 40 % and 42 %, respectively.

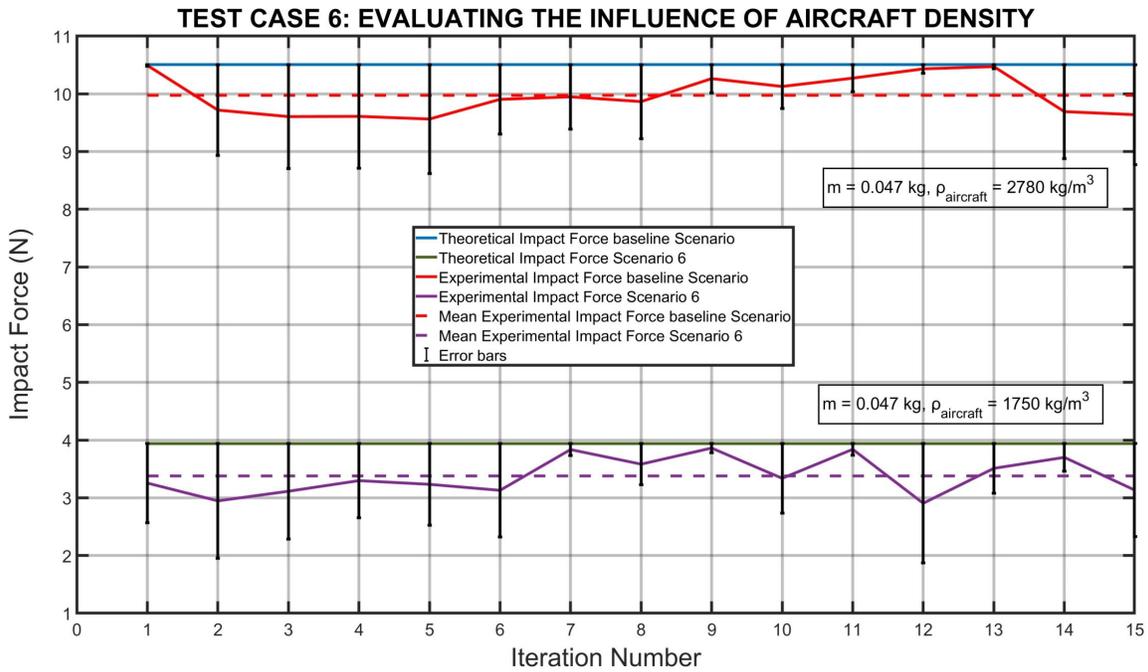

Fig. 15: Influence of aircraft density (Test Case 6)

Figure 15 illustrates the influence of aircraft density. By using CFRP as a test specimen, which has 58 % less density compared to Aluminum, the impact force theoretically decreases by 62 %. In the bird strike experiments, the average reduction in impact force was measured to be 65 %.

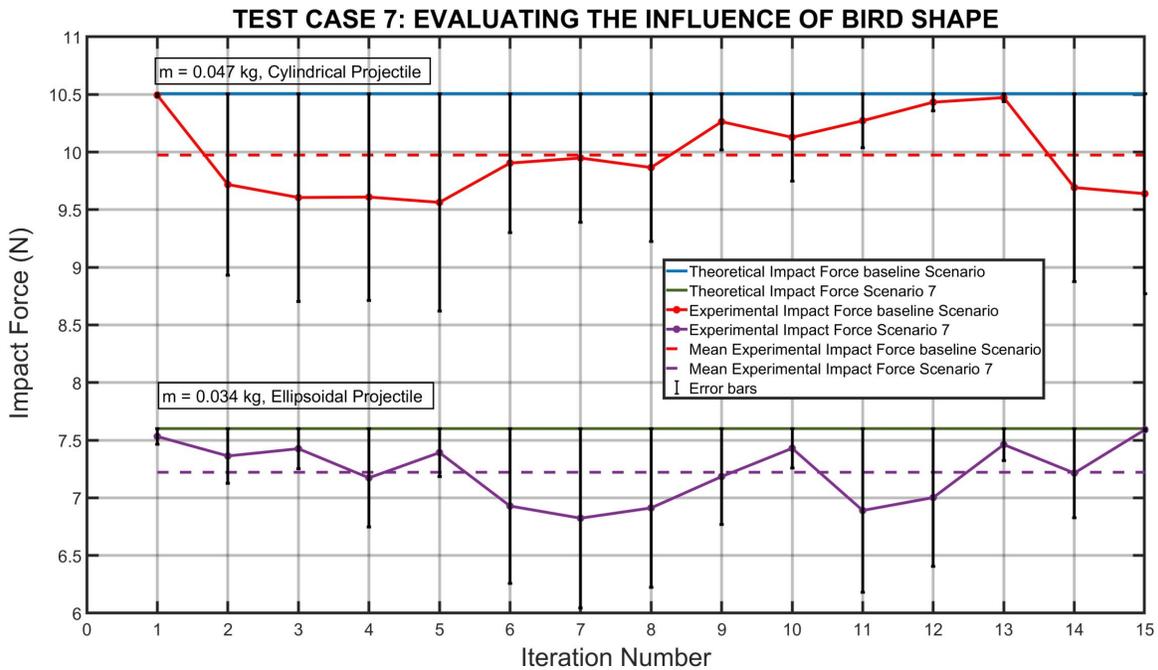

Fig. 16: Influence of bird shape (Test Case 7)

The effect of bird shape is shown in Figure 16. The results demonstrate that the ellipsoidal projectile exerts 27 % less impact force than the cylindrical projectile in theory primarily because of its lower mass and rounded shape. In the practical experiments, the mean reduction in impact force was 35 %.

Eventually, Figure 17 illustrates mean percentage conformance and standard deviation of the experimental results obtained in individual test scenarios with the theoretical impact force model expressed in Equation 6. The mean percentage conformance is explained in the Equation 16.

$$\% \text{ error} = \left[\frac{(\text{theoretical results - experimental results}) \cdot 100}{\text{theoretical results}}\right] \quad (15)$$

$$\text{mean \% conformance} = 100 - \text{mean \% error} \quad (16)$$

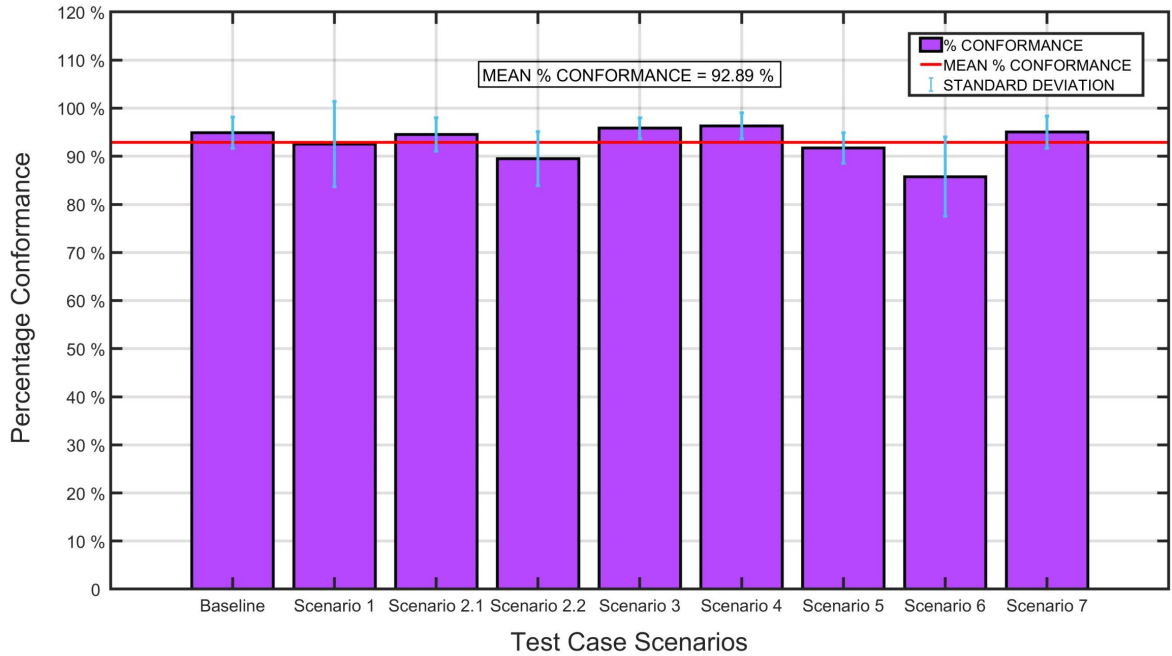

Fig. 17: Percentage Conformance of the experimental results with the theoretical estimation

It can be observed that the scenario 4 shows the maximum conformance of 96 % while the scenario 6 showed the minimum conformance of 85 %. The mean percentage conformance of all the test cases was 92 %. The next section discusses the interpretation, analysis and explanation of the obtained experimental results.

## 5 DISCUSSION

In order to have an enhanced insight of the bird strike problem in the context of UAM, a theoretical impact force model considering the underlying factors of bird strike was developed [7] which quantified the exerted impact force and kinetic energy due to the collision. The goal of this research was to validate this theoretical impact force model by performing impact force experiments and to compare the results to the outcomes of the theoretical model. Identical to the theoretical impact force model, the experimental model developed in this research also quantifies the generated impact force due to the bird strike for different test cases. In this paper, seven test cases are formulated for conducting the bird strike experiment.

In all the test cases, the conformance exceeded 85 %. Notably, the lowest % conformance equal to 86 % was observed in test case 6. Assessing this in terms of impact force measurements, this attributed to an average difference of 0.7 N between the theoretical predictions and experimental measurements. Similarly, corresponding to the highest % conformance of 96 % for test case 4, the average difference between theory and experiments amounted to 0.3 N. According to the certification requirements, an air taxi should withstand a maximum impact force of 2255 N for a single bird strike 4819 N for flocks [7], [8]. Therefore, considering the substantial magnitudes of impact force involved in bird strikes, the error in impact force measurements below 1 N demonstrates a high level of accuracy. Moreover, it was observed in all the test cases that magnitude of experimental impact force was on an average 7 % less than the theoretically predicted values. This difference can be attributed to inaccuracies in force sensor measurements, error in impact velocity measurements and the angle of impact not being equal to 90° for every iteration. From the impact force Equation 6, it can be seen that if the collision was not head-on or $\theta$ was not equal to 90°, the impact force reduces. As it was difficult to practically reproduce exact head-on collisions, the experimental impact force values are always less than the theoretical baseline. The fluctuations between the individual iterations per test case are due to precision of measuring instruments used in the experiment, inconsistencies and lack of repeatability due to human involvement as the bird projectiles were dropped manually by hand and challenges in achieving high surface quality and resolution in 3D printed projectiles leading to variation in characteristic properties such as mass and density. Moreover, load cells may exhibit non-linear behaviour, meaning that the relationship between the applied load and the electrical output signal may not be perfectly linear. Load cells are also sensitive to electrical noise, temperature, humidity and mounting angle [20].

The experimental setup has the potential for improvement by eliminating human involvement and implementing a mechanism for dropping the projectile to perform it consistently for all experimental iterations. This modification would enhance the experiment's repeatability as well as ensure an accurate impact angle. Additionally, utilizing a guided launching system for the bird projectile can ensure consistent impact spots throughout all iterations. This approach would increase the accuracy of force measurements. Another alternative to achieve precise force measurement is the utilization of multiple strain gauges instead of relying on a single load cell. However, the current setup did not incorporate strain gauges due to their complexities in mounting and assembly.

In addition to this, it was found during the design of the experiment that the drop-weight mechanism was not able to produce full scale impact velocities due to the involved physics and spatial constraints of the test facility in achieving the required drop-height. Therefore, the impact velocities were scaled down by the factor of 1:15 in order to attain a feasible range of drop-height. The consequences of scaling down the impact velocity are discussed below.

The impact force in the current experimental analysis with scaled down impact velocity was marked down to be 0.4 % of the full scale impact force. Therefore, the range of scaled down values affected the choice of impact force sensor and its required measurement range for the experiment. Additionally by scaling down the impact velocity, the parasitic drag on the projectile decreased as it is directly proportional to the square of impact velocity and the skin friction drag also reduced because of decrease in Reynolds Number and geometric dimensions of the projectile [13]. Therefore, the difference between the actual impact velocity that was measured during the experiment and the approximated impact velocity $v = \sqrt{2 \cdot g \cdot h}$ which was used for calculating the drop-heights was minimal. In addition to that, after scaling down the impact velocity, the influence of wind was also minimal as the experiment can be carried out in close quarters of the laboratory. Eventually, the choice of drop-weight mechanism and scaling down the impact velocity can be justified with the fact that it was not necessary to reproduce the actual aircraft speeds and bird speeds as the test specimen in this research was not subjected to certification but the goal was to validate the impact force model which in turn can be used to propose recommendations on the current certification requirements. Overall, the theoretical impact force calculations for every test case are compliant with the acquired experimental impact force data and the theoretical impact force model determined in the paper [7] is valid for quantification of impact force for bird strikes in the UAM architecture.

# 6 CONCLUSION

In this paper, an experimental model representing a bird strike in the UAM architecture was developed to validate a theoretical impact force model. For the design and development of the experimental model, it was divided into several aspects namely the launching mechanism, the bird projectile, the test specimen, the sensors and data acquisition system. Different alternatives were investigated for each aspect and these alternatives were evaluated against the high level technical and operational requirements of the experiment. The best suitable alternatives satisfying all the requirements were selected for the final development of the model. Subsequently, a test matrix containing seven test cases, nine test scenarios and 135 iterations was formulated to conduct the bird strike experiment and to validate the theoretical framework for variation in the underlying parameters of collision. The influencing parameters namely bird mass, bird speed, aircraft speed, bird density, bird length, aircraft density, angle of impact and bird shape were considered for theoretical model verification. The experimental results showed an average conformance of 92 % and the average difference between theoretical predictions and experimental values was 0.5 N. Based on the obtained results for all the test cases, it can be concluded that the theoretical computations are valid for changes in the influencing parameters of a bird strike. Thus, the theoretical impact force model is validated for all the test cases and test scenarios. The validation of the model further implies that the theoretical framework emerges as the possible solution to evaluate and quantify the consequences of collision between air taxis and birds in terms of generated kinetic energy and impact force.

However, there are certain limitations of the experimental setup such as human involvement for dropping the projectile, absence of guided impact, electrical and environmental interference in load cells and low surface quality and resolution of 3D printed projectiles. Additionally, the two influencing parameters which are not considered in the experimental model are depth of penetration and surface curvature. Future work will eliminate the human involvement as well as include a guided impact of the projectile to increase accuracy in force measurements and repeatability. In addition to this, a method will be devised to quantify the depth of penetration in the bird strike experiment and accounting for surface curvature in the theoretical estimation of impact force. Thereafter, the effect of surface curvature will be validated experimentally. This may allow for a more accurate estimation of impact force and potential enhancement of the theoretical framework.

## Acknowledgment

I would like to express my deepest gratitude to all those who have supported and contributed to the completion of this research paper. I would also like to acknowledge the support provided by Assistant Professorship of eAviation, Technical

6 CONCLUSION                                                                                                                                              20University of Munich where this research was conducted. The resources and facilities made available by the institution were essential in conducting experiments, analyzing data, and producing meaningful results. Finally, I would like to express my gratitude to my family for their unwavering support, encouragement, and understanding throughout this research endeavor. Their patience, motivation, and belief in my abilities have been a constant source of inspiration.

# References

[1] C. Silva, W. R. Johnson, E. Solis, M. D. Patterson, and K. R. Antcliff, "VTOL urban air mobility concept vehicles for technology development," in *2018 Aviation Technology, Integration, and Operations Conference*, 2018, p. 3847.

[2] Federal Aviation Administration and United States Department of Transportation, "Urban Air Mobility and Advanced Air Mobility," 2021. [Online]. Available: https://www.faa.gov/uas/advanced_operations/urban_air_mobility/

[3] Deloitte, "National Aeronautics and Space Administration (NASA) UAM Vision Concept of Operations (ConOps) UAM Maturity Level (UML) 4," 2020.

[4] R. A. Dolbeer, M. J. Begier, P. R. Miller, J. R. Weller, A. L. Anderson *et al.*, "Wildlife strikes to civil aircraft in the united states, 1990–2021," United States. Department of Transportation. Federal Aviation Administration, Tech. Rep., 2021. [Online]. Available: https://www.faa.gov/airports/airport_safety/wildlife/wildlife_strikes_civil_aircraft_united_states_1990_2021

[5] Statista, "Cruising speeds of the most common types of commercial airliners." 2016, https://www.statista.com/statistics/614178/cruising-speed-of-most-common-airliners/. Online: accessed September 30 2016.

[6] R. Goyal, C. Reiche, C. Fernando, J. Serrao, S. Kimmel, A. Cohen, and S. Shaheen, "Urban air mobility (UAM) market study," Tech. Rep., 2018. [Online]. Available: https://ntrs.nasa.gov/api/citations/20190001472/downloads/20190001472.pdf

[7] A. Devta, I. C. Metz and S. F. Armanini, "Evaluation And Quantification of the Potential Consequences of Bird Strikes in Urban Air Mobility," *33rd ICAS 2022*, 2022.

[8] European Union Aviation Safety Agency (EASA), "Third Publication of Proposed Means of Compliance with the Special Condition VTOL MOC-3 SC-VTOL," 2022. [Online]. Available: https://www.easa.europa.eu/en/document-library/product-certification-consultations/special-condition-vtol

[9] F. Beer, E. Johnston, J. DeWolf, and D. Mazurek, "Mechanics of materials. 7th_edition," *New York. MeGraw-Hill Education Ltd*, 2015.

[10] B. Bruderer and A. Boldt, "Flight characteristics of birds: I. radar measurements of speeds," *Ibis*, vol. 143, no. 2, pp. 178–204, 2001.

[11] D. M. Hamershock, T. W. Seamans, and G. E. Bernhardt, "Determination of body density for twelve bird species," WRIGHT LAB WRIGHT-PATTERSON AFB OH, Tech. Rep., 1993.

[12] B. MacKinnon, R. Sowden, and S. Dudley, "Sharing the skies: an aviation guide to the management of wildlife hazards. transport canada," 2001.

[13] P. M. Whelan and M. J. Hodgson, *Essential Principles Of Physics*. London, 1987.

[14] NASA Glenn Research Center, "Terminal velocity," 2009. [Online]. Available: https://www.grc.nasa.gov/www/k-12/airplane/termv.html

[15] J. S. Wilbeck and J. L. Rand, "The development of a substitute bird model," 1981.

[16] J. Riegel, W. Mayer, and Y. van Havre, "Freecad," 2016.

[17] A. Akash, V. S. J. Raj, R. Sushmitha, B. Prateek, S. Aditya, and V. M. Sreehari, "Design and Analysis of VTOL Operated Intercity Electrical Vehicle for Urban Air Mobility," *Electronics*, vol. 11, no. 1, p. 20, 2021.

[18] T. H. G. Megson, *Aircraft structures for engineering students*. Butterworth-Heinemann, 2016.

[19] S. Al-Mutlaq, "Getting started with load cells," *Conteúdo Online*, 2016. [Online]. Available: http://tet.pub.ro/Documente/Proiect%20final/Documentatie/Magnetormetru%20MAG3110/Getting%20Started%20with%20Load%20Cells%20-%20learn.sparkfun.pdf

[20] S. Al-Mutlaq and A. Wende, "Load cell amplifier hx711 breakout hookup guide," *Retrieved from Sparkfun Start Something*, 2016. [Online]. Available: https://learn.sparkfun.com/tutorials/load-cell-amplifier-hx711-breakout-hookupguide/introduction

[21] M. Banzi and M. Shiloh, *Getting started with Arduino*. Maker Media, Inc., 2022.

[22] W. Christian, M. Belloni, F. Esquembre, B. A. Mason, L. Barbato, and M. Riggsbee, "The physlet approach to simulation design," *The Physics Teacher*, vol. 53, no. 7, pp. 419–422, 2015.